\documentclass{article}

\usepackage{arxiv}

\usepackage[utf8]{inputenc} % allow utf-8 input
\usepackage[T1]{fontenc}    % use 8-bit T1 fonts
\usepackage{hyperref}       % hyperlinks
\usepackage{url}            % simple URL typesetting
\usepackage{booktabs}       % professional-quality tables
\usepackage{amsfonts}       % blackboard math symbols
\usepackage{nicefrac}       % compact symbols for 1/2, etc.
\usepackage{microtype}      % microtypography
\usepackage{cleveref}       % smart cross-referencing
\usepackage{lipsum}         % Can be removed after putting your text content
\usepackage{graphicx}
\usepackage{natbib}
\usepackage{doi}

\DeclareMathSymbol{\blacktriangle} {\mathord}{AMSa}{"4E}
\DeclareMathSymbol{\blacksquare} {\mathord}{AMSa}{"04}
\DeclareMathSymbol{\blacklozenge} {\mathord}{AMSa}{"07}

\title{Diversity in Axelrod's social adaptation model}

\newif\ifuniqueAffiliation

\uniqueAffiliationtrue

\ifuniqueAffiliation % Standard variant of author block

\author{ 
{\hspace{1mm}Juan ~Villegas-Febres}\thanks{Also: juanvillegas.febres@gmail.com. Paper originally published in spanish in UNET Scientific Journal. Volume 20(2):2008 (2009). ISSN 1316-869X.} \\
	Grupo de Qu\'imica Te\'orica QUIFFIS\\
	Facultad de Ciencias\\
	Universidad de Los Andes \\
	M\'erida 5101\\
	Venezuela\\
	\texttt{juancv@ula.ve} \\
	\And
	{\hspace{1mm}Emma ~Castillo-Felisola}\\
	Grupo de Qu\'imica Te\'orica QUIFFIS\\
	Facultad de Ciencias\\
	Universidad de Los Andes \\
	M\'erida 5101\\
	Venezuela\\
	\texttt{castillofelisolaemmabeatriz@gmail.com}\\}

\else

\usepackage{authblk}

\author[1]{

	\hspace{1mm}Juan C. Villegas-Febres\thanks{\texttt{juancv@ula.ve, castillofelisolaemmabeatriz@gmail.com}}}
}

\affil[1]{Grupo de Qu\'imica Te\'orica: Quimicof\'isica de Fluidos y Fen\'omenos Interfaciales (QUIFFIS), Facultad de Ciencias, Universidad de Los Andes, M\'erida 5101, Venezuela}

\fi

\hypersetup{
pdftitle={Diversity in Axelrod's social adaptation model},
pdfauthor={Juan ~Villegas-Febres, Emma ~Castillo-Felisola},
pdfkeywords={Axelrod's model, cultural diversity, cellular automata, social temperature},
}

\begin{document}
\maketitle

%\e^{

\begin{abstract}
%	\lipsum[1]
In this paper, we study Axelrod's model of social dynamics, introducing the concept of {\it Cultural Diversity} ($D$), defined as the variety of sizes of {\it clusters} or cultural domains formed, which measures the complexity of the system. We find that the maximum of $D$ agrees with the critical point where the {\it monocultural/multicultural} phase transition occurs, tending to a minimum value when the system's degrees of freedom, cultural traits $q$, are far from the critical point $q_c$. We show that at $q_c$ the entropy also reaches its maximum value, that is, the phase transition for this model is of the {\it order-order} type. Thus, {\it multiculturalism} is not synonymous with cultural diversity, as is commonly assumed in the literature.
\end{abstract}


%xxxxxxxxxxxxxxxxxxxxxxxxxxxxxxxxxxxxxxxxxxxxxxxxxxxxxxxxx
\section{Introduction}
%xxxxxxxxxxxxxxxxxxxxxxxxxxxxxxxxxxxxxxxxxxxxxxxxxxxxxxxxx

%\cite{kour2014fast}
Since the equivalence between entropy and information transmission was demonstrated \cite{Shannon-1949}, a formal bridge between the description of physical systems and society has been made possible. However, a serious difficulty that has always existed between scientists and philosophers is mutual distrust; the former have shown little interest in the rich philosophical and sociological tradition accumulated over the last two thousand years, and the latter have shown a kind of prejudice that prevents the mere acceptance of the possibility of studying society and its dynamics using methods developed in the physical and mathematical sciences. 

In recent years, in the context of the study of complex systems, and using tools and methods developed in thermodynamics and statistical mechanics, there has been a serious and massive effort to scientifically elucidate some of the mechanisms underlying a series of complex economic and social phenomena of great interest: opinion formation, self-organization, wealth distribution, coalition formation, land and air traffic, information networks, etc.

Within these phenomena and systems, the model of social adaptation developed by Robert Axelrod \cite{Axelrod-1997} deserves particular attention. It attempts to answer the following question \cite{Axelrod-2004}: 

{\it "If people tend to become more similar in their beliefs, attitudes, and behaviors when they interact, why don't all their differences disappear at some point?"} 

He found interesting and counterintuitive results, given the existence of a threshold beyond which an {\it order-disorder} transition becomes evident, that is, a "phase shift": {\it monocultural/multicultural}. This type of phenomenon implies the existence of collective, cooperative effects that also appear in many physical and chemical systems and processes. 

Using the Axelrod model, in one (1D) and two (2D) dimensions, cultural {\it drift}
\cite{Mazzi-2007,Klemm-2003b}
%[Mazzitello et al. 2007, Klemm et al. 2003b] 
and transmission \cite{Klemm-2003a, Castellano-2000, Vilone-2002}
%[Klemm et al. 2003a, Castellano et al. 2000, Vilone et al. 2002] 
have been studied, as well as the effect of propaganda on a model society 
\cite{Gonzalez-2005, Gonzalez-2006, Mazzi-2007}
%[González et al. 2005, González et al. 2006, Mazzitello et al. 2007]. 

Recently, the absolute temperature $T^{*} = 
(\partial \overline{E} / \partial \overline {S})$ was calculated \cite{Villegas-2008}
%[Villegas et al. 2008] 
for the Axelrod model, using the Lyapunov potential ($L$), which is closely related to the agent-reduced energy ($\overline{E}$) and the agent-reduced Boltzmann-Gibbs-Shannon entropy ($\overline{S}$). It was also shown that this model exhibits negative absolute temperatures, which is observed in real quantum systems with magnetic spins \cite{Purcell, Ramsey}
%[Purcell et al. 1951, Ramsey 1956] 
and in no way contradicts the second law of thermodynamics \cite{Landau, kikoin-1979}.
%[Landau et al. 1980, Kikoin et al. [1979]. 
Villegas-Febres et al.(\cite{Villegas-2008}) found that the order-disorder phase transition (monocultural/multicultural) occurred precisely where the temperature changed sign, that is, where $\overline{S}$ was maximum ($\overline{S}_c$).

The prerequisite for the existence of these temperatures is that the system has a limited number of energy states, and therefore a maximum value \cite{Mosk-2005}. An increase in energy leads to an increase in the accessible states (entropy) of the system. The temperature in this case is "normally" positive. As the energy continues to increase, the system is distributed uniformly over all its possible energy values, thus obtaining a maximum entropy. This situation therefore corresponds to an infinitely high temperature \cite{kikoin-1979}. An even greater increase in energy invariably leads to a {\it population inversion} (which is the basis, for example, of laser amplification) and therefore to a decrease in entropy, due to the gradual "accumulation" of the particles that make up the system at their highest possible energy level. This then gives rise to the existence of negative temperatures. A negative temperature means that it is higher than an "infinitely high" temperature \cite{kikoin-1979}. 

When Tsang \cite{Tsang-2000} studied the percolation phenomenon using a two-dimensional square network with periodic boundary conditions, he also found that $\overline{S}$ displayed a maximum when the system "percolated." He used a checkerboard-like array of squares, each with probability $p$ of being "occupied" and probability $1 - p$ of being "unoccupied." The system is said to "percolate" to a critical probability value $p = p_c$ when a group of occupied neighboring squares, forming a cluster, extends from one end of the two-dimensional network to the other. Using the Hoshen-Kopelman algorithm
\cite{Hoshen}, Tsang (\cite{Tsang-2000}) was able to determine the {\it Cluster Diversity} $D$; understood as the number of clusters of different sizes formed in the network, for each value of $p$. He found that the maximum diversity, $D_c$, occurs exactly at the critical point where the entropy $\overline{S}$ is also maximum, $\overline{S}_c$. Tsang used $D$ as a measure of system complexity. In the context of the Axelrod model, $D$ measures the distinct number of cultural groups ({\it clusters} of neighboring elements that share the same cultural vector $C_i$) formed by one, two, three, etc. agents.

In this work, we combine the results obtained by Villegas-Febres et al. (\cite{Villegas-2008}) regarding the use of entropy and temperature as {\it order parameters}, as well as the diversity calculations performed by Tsang (\cite{Tsang-2000}) in percolation networks. This leads to a reinterpretation of the monocultural/multicultural transition as an order-to-order transition, rather than the typical order-to-disorder transition, and also clarifies the difference between multiculturalism (as a synonym for "cultural richness") and cultural diversity.

%xxxxxxxxxxxxxxxxxxxxxxxxxxxxxxxxxxxxxxxxxxxxxxxxxxxxxxxxx
\section{Method}
%xxxxxxxxxxxxxxxxxxxxxxxxxxxxxxxxxxxxxxxxxxxxxxxxxxxxxxxxx

The Axelrod social adaptation model, which we use in this work, consists of $N = n^2$ agents located in a two-dimensional $n \times n$ square network, where each agent interacts with its four nearest neighbors, with periodic boundary conditions. The cultural state $C_i$ of each agent $i$ is defined by a vector of $F$ components, $C_i = (\sigma_{i1} , \sigma_{i2} , . . . , \sigma_{iF} )$, which identifies its {\it attributes}. Each of the {\it traits} $\sigma_{if}$ if can, in turn, take any of the $q$ positive integer values in the interval $[1, q]$. There are, therefore, $q^F$ possible cultural vectors accessible to each of the agents in this model social system.

The dynamic simulation process, given specific values of $N$, $F$, and $q$, follows the following sequence of steps:

(1) Initially, a culture vector $C_i$ is randomly generated for each agent $i$.

(2) An agent $i$ is randomly selected from the network, whose culture vector is $C_i$.

(3) The overlap (number of traits $\sigma_{if}$ if they share) $l(i,j) = \sum_{f = 1}^{F} \delta_{\sigma_{if},\sigma_{jf}}$, between agent $i$ and one of its four neighbors $j$, also chosen at random, is calculated;
where $\delta_{\sigma_{if},\sigma_{jf}} = 1$, if 
$\sigma_{ik} = \sigma_{jk}$, and $\delta_{\sigma_{if},\sigma_{jf}} = 0$, if 
$\sigma_{ik} \neq \sigma_{jk}$.

(4) If $0 < l (i, j) < F$, agent $i$ interacts with neighbor $j$ with probability $p_{ij} = l(i,j)/F$. In this case, a feature $k$ not common to both agents is chosen, so that initially $\sigma_{ik} \neq \sigma_{jk}$ must hold.
Then, both features are equalized, $\sigma_{ik} = \sigma_{jk}$, provided that $p_{ij} \geq \zeta$, where $\zeta$ is a randomly generated number, with $0 \leq \zeta \leq 1$.

The above process is repeated from $(2)$ to $(4)$ until a {\it frozen} or {\it absorbing} situation is obtained, where, for each pair of neighboring agents $ij$ in the system, one of these two possibilities holds:
$l(i,j) = 0$ or $l(i,j) = F$. The sequence from $(2)$ to $(4)$ required, for each value of $q$, between $10^8$ and $10^9$ cycles. 
This set of cycles is then repeated again, from $(1)$ to $(4)$, $n_s$ times. The values reported in this work result from taking the average over $n_s$ independent runs, shown between the symbols $< >$. Two clearly differentiated situations then arise: homogeneous or {\it monocultural}, where $\sum_{i<j} l(ij) = 2 N F$, and heterogeneous or {\it multicultural}, where $0 \leq \sum_{i<j} l(ij) < 2 N F$. The term $2 N F$ is the maximum total number of links that can occur in a two-dimensional network, with $4$ neighbors per agent. In the first case, the entire system is in a single cultural state $C_i$, in any of its possible $q^F$ values. In the second case, two or more cultural {\it domains} appear.

A domain (or {\it cluster}) is a set of contiguous elements (agents) that share the same cultural vector $C_i$. Domains, in turn, are connected to other domains by means of elements with zero overlap.

It has been shown that Axelrod's social adaptation model, in one and two dimensions, shows a monocultural/multicultural transition for a specific value of $q = q_c$, for a given value of $N$ and $F$. Thus, when $q < q_c$, the system tends to be monocultural, and when $q > q_c$, the system tends to be multicultural. When the number of agents in the cultural system tends to be very large ($N \rightarrow \infty$), this transition is much better defined (\cite{Castellano-2000,Klemm-2003b}),
%[Castellano et al. 2000, Klemm et al. 2003a], 
resembling a typical first-order transition. The Axelrod model shows that, although the probability $p_{ij}$ of overlap increases as neighboring agents $i$ and $j$ become increasingly similar, above a certain critical value $q_c$, this leads to a surprising result of multiculturalism. 

Characterizing the critical point, and the reasons why it occurs, is essential. In the literature \cite{Castellano-2000,Gonzalez-2005}
%[Castellano et al. 2000, González et al. 2005], 
two criteria have typically been used as {\it order parameter} in the absorptive state: a) The average relative size of the largest cultural domain reached, $\langle M_{MAX}\rangle / N $; and 
b) The fraction of average cultural domains, $\langle N_D \rangle / N$, where $N_D$ is the number of domains formed.
For the monocultural state, it holds that $\langle M_{MAX}\rangle / N \rightarrow 1$ and $\langle N_D\rangle /N \rightarrow 0$, while for the multicultural state, 
$\langle M_{MAX} \rangle /N \rightarrow 0$ and $\langle N_D\rangle / N \rightarrow 1$. All of this is more evident as 
$N \rightarrow \infty$.

Based on Tsang's studies (\cite{Tsang-2000}), we calculate cultural diversity as $D = \sum_{s} \Theta [N (s,p)]$, where $N(s,q)$ represents the number of domains of size $s$ formed, for a given value of $q$, and $\Theta (x)$ is a {\it Heaviside function}, where

a) $\Theta (x) = 1, \forall x > 0$;

b) $\Theta (x) = 0, \forall x = 0$.

As already mentioned, the Hoshen-Kopelman algorithm, developed in the study of percolation phenomena and {\it cluster} enumeration, allows the determination of the number and size of domains formed in the simulation process.

%xxxxxxxxxxxxxxxxxxxxxxxxxxxxxxxxxxxxxxxxxxxxxxxxxxxxxxxxx
\section{Results}
%xxxxxxxxxxxxxxxxxxxxxxxxxxxxxxxxxxxxxxxxxxxxxxxxxxxxxxxxx

%111111111111111111111111111111111111111111111111111
\begin{figure}[h! tbp]
	\centering
%	\fbox{\rule[-.5cm]{4cm}{4cm} \rule[-.5cm]{4cm}{0cm}}
%	\caption{Sample figure caption.}
 %\includegraphics[width=\linewidth]
 \includegraphics[width=12.7cm]
    {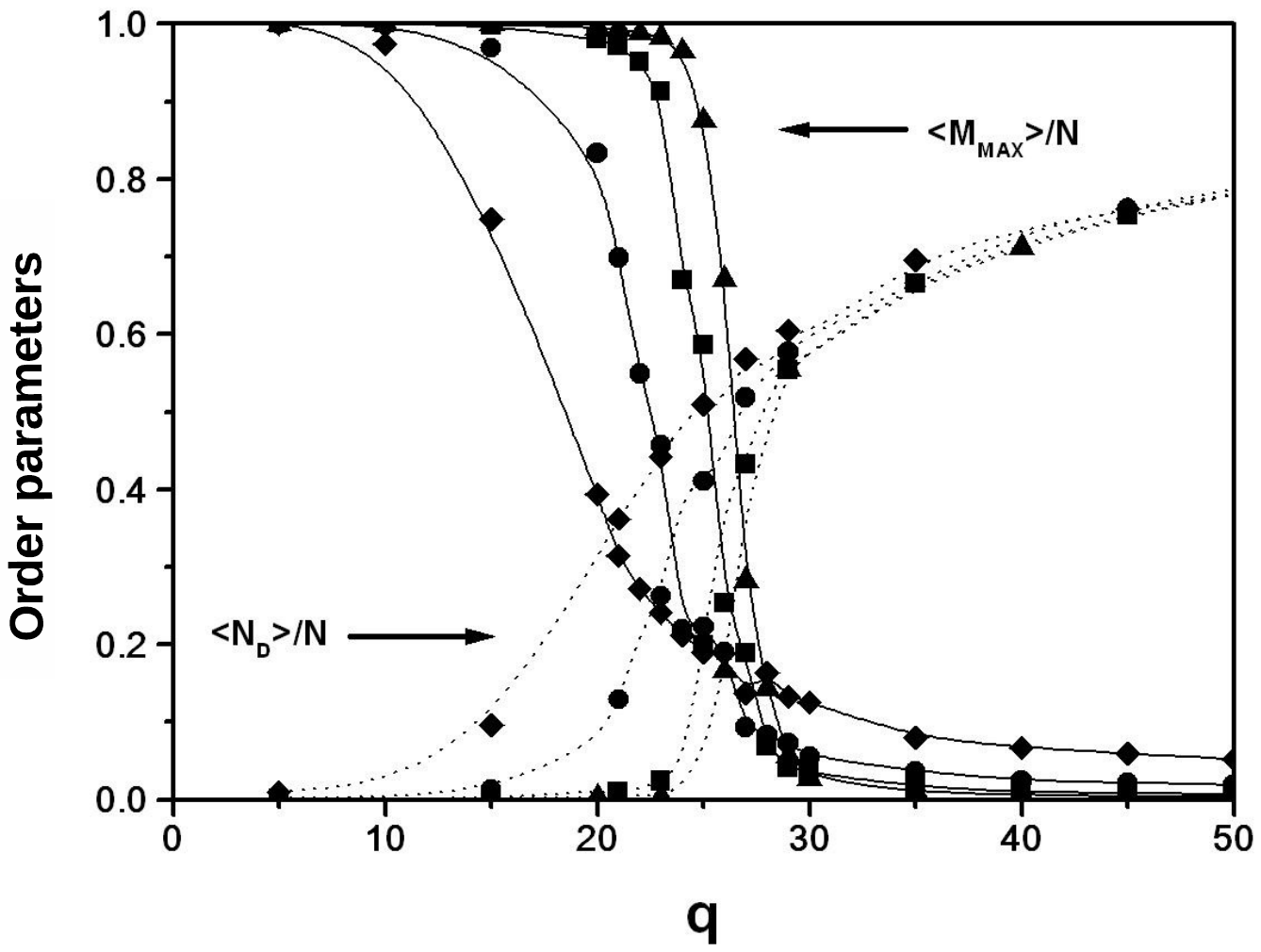}
 \caption{Order parameters, $\langle M_{MAX} \rangle/N$ (solid line) and $\langle N_D\rangle /N$ (dotted line), as a function of $q$, for $N = 10 \times 10$ (diamonds), 
 $N = 20 \times 20$ (circles), $N = 40 \times 40$ (squares), and $N = 60 \times 60$ (triangles). The solid and dotted lines represent the curves that best fit the simulation results. For all cases, $F = 5$.}
 %.
 \label{figure1}
 %    }
%
\end{figure}
%111111111111111111111111111111111111111111111111111

Figure 1 shows the results for the order parameters $\langle M_{MAX}\rangle / N$ and $\langle N_D\rangle / N$ as a function of $q$, for $N = 10 \times 10$, $N = 20 \times 20$, $N = 40 \times 40$, and $N = 60 \times 60$, represented in the figure by rhombuses, circles, squares, and triangles, respectively, where $F = 5$ for all cases. The solid and dotted lines represent the curves that best fit the simulation results for $\langle M_{MAX}\rangle / N$ 
and $\langle N_D\rangle / N$ , respectively. The results were averaged over $n_s = 50$ independent runs, under different initial conditions of $C_i$ obtained at random. For conditions close to the critical point, $n_s = 100$. The numerical results throughout this work will always be under these same conditions. It is observed that as $N$ increases, the results tend asymptotically toward a limiting value. These results are similar to those obtained by González (\cite{Gonzalez-2006}), for the case when the external field $B$ (which simulates the effect of propaganda on social agents) is equal to zero and $N = 40 \times 40$, where $q_c \cong 25$. In recent work \cite{Villegas-2008}, we had already reported these results are true for $N = 20 \times 20$ and $N = 40 \times 40$.

%222222222222222222222222222222222222222222222222222
\begin{figure}[h! tbp]
	\centering
%	\fbox{\rule[-.5cm]{4cm}{4cm} \rule[-.5cm]{4cm}{0cm}}
%	\caption{Sample figure caption.}
% \includegraphics[width=\linewidth]
 \includegraphics[width=12.7cm]
    {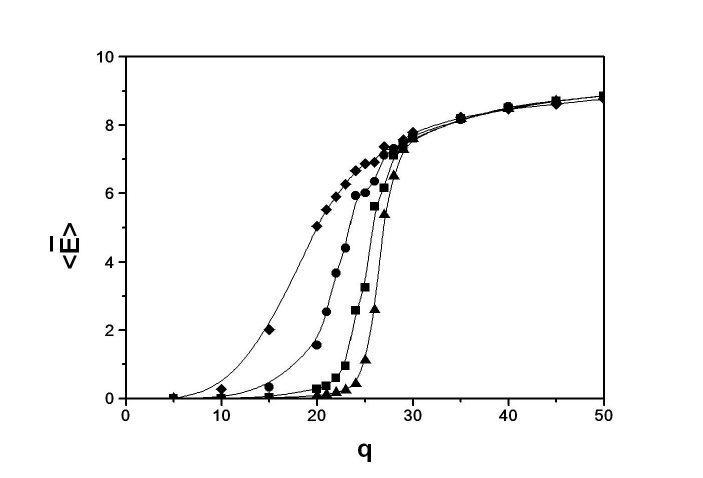}
 \caption{Average energy reduced per agent $\langle \overline E\rangle$ as a function of $q$.}
 %.
 \label{figure2}
 %    }
%
\end{figure}
%222222222222222222222222222222222222222222222222222

A very important aspect is to study the stability of absorbing states. If we define the Lypanuov potential (\cite{Klemm-2003a,Klemm-2003b})
%[Klemm et al. 2003a, Klemm et al. 2003B] 
as the negative of the total number of {\it overlapping bonds} $(l(ij) = 1, \forall \,ij) L = - \sum_{i<j} l(ij)$, and $E$ as the total number of 
{\it non-overlapping bonds} $(l(ij) = 0, \forall \,ij)$, then
$E - L = 2 N F$. In the homogeneous case (monocultural), $L = L_0 = - 2 N F$, and therefore $E = E_0 = 0$. In the heterogeneous case (multicultural), $L \rightarrow 0$ and $E \rightarrow 2 N F$. It is clear then that the energy reduced by the agent $\overline E = E/N \rightarrow 0$ for the homogeneous state, and $\overline E = E/N \rightarrow 2 F = 10$ for the inhomogeneous state.

Figure 2 shows the average results of reduced $E$ per agent, $\langle \overline E \rangle$, as a function of $q$. The symbology and conditions are the same as in Figure 1. It is worth noting that $\langle \overline E \rangle$ always increases as cultural traits $q$ increase. If the traits shared by agents are a measure of the system's stability, then $\overline E$ represents a direct measure of its energy. 

%33333333333333333333333333333333333333333333333333333
\begin{figure}
	\centering
%	\fbox{\rule[-.5cm]{4cm}{4cm} \rule[-.5cm]{4cm}{0cm}}
%	\caption{Sample figure caption.}
% \includegraphics[width=\linewidth]
 \includegraphics[width=12.7cm]
    {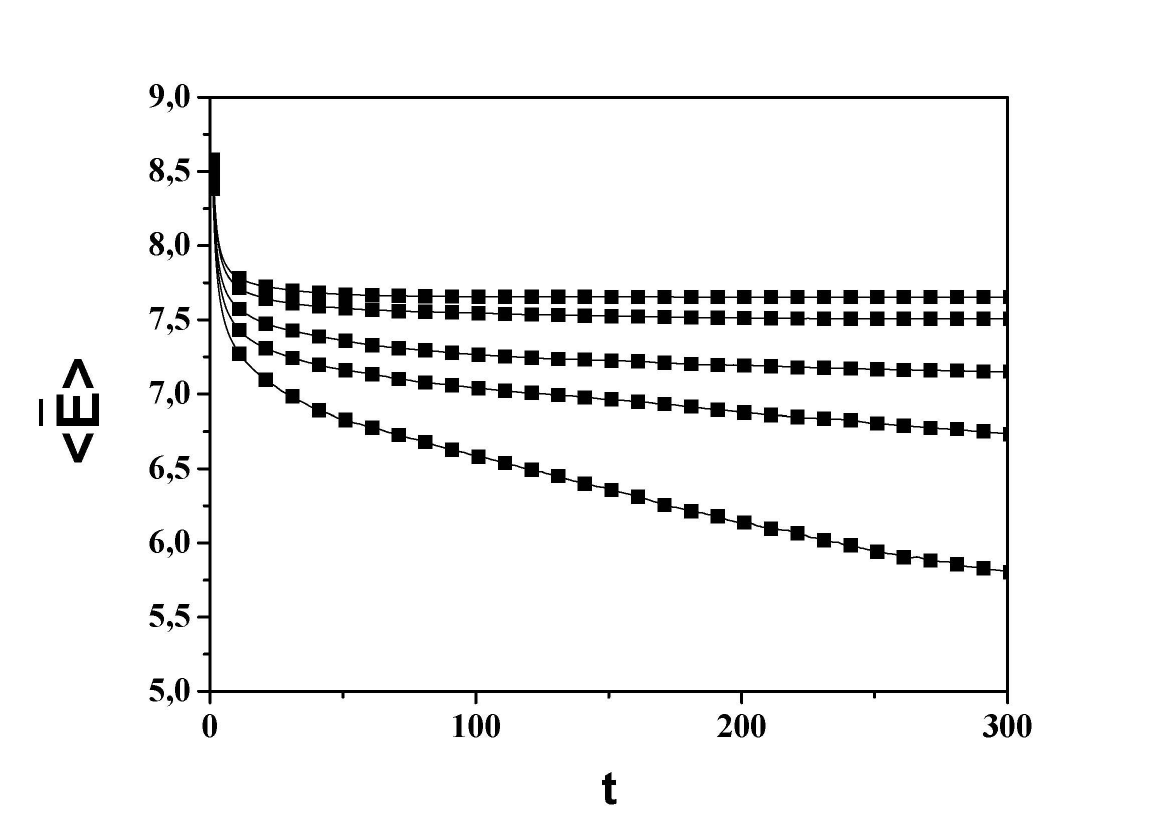}
 \caption{Average energy reduced by agent $\langle \overline E\rangle$ as a function of time $t$, for $N = 40 \times 40$. From bottom to top, from $q = 26$ to $q = 30$.}
 %.
 \label{figure3}
 %    }
%
\end{figure}
%33333333333333333333333333333333333333333333333333333

Figure 3 presents the dynamic results of $\langle \overline E \rangle$ as a function of t for the $N = 40 \times 40$ system. Five values are shown, above and below the critical point: $q = 26$, $q = 27$, $q = 28$, $q = 29$, and $q = 30$. Each unit of "time" ($t$) represents $10^5$ simulation steps described above. For each value of $t$, the average of $50$ independent runs was taken. It is clear that statistically, the energy always decreases with time; therefore, we can associate it with a Lyapunov-type potential, which always decreases dynamically.
\cite{Klemm-2003c}
%[Klemm et al. 2003c] 
had already shown that in the case of the one-dimensional Axelrod model, the negative of the total overlaps $L(t)$ was indeed a Lyapunov potential, since it always decreased during the evolutionary dynamics. In the case of our work, we have shown numerically that $L$, and therefore the energy $\overline E$, are indeed Lyapunov-type potentials, since the difference between these two quantities is a constant equal to $2 F$ .

%444444444444444444444444444444444444444444444444444
\begin{figure}
	\centering
%	\fbox{\rule[-.5cm]{4cm}{4cm} \rule[-.5cm]{4cm}{0cm}}
%	\caption{Sample figure caption.}
% \includegraphics[width=\linewidth]
 \includegraphics[width=12.7cm]
    {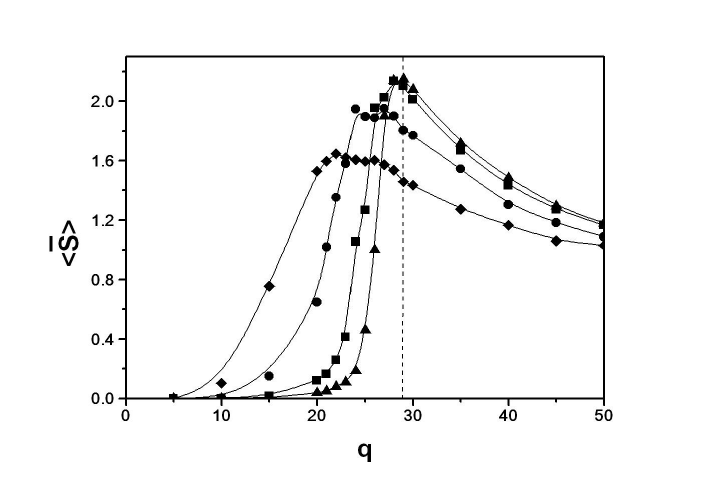}
 \caption{Average reduced entropy per agent $\langle \overline S\rangle$ as a function of $q$.}
 %.
 \label{figure4}
 %    }
%
\end{figure}
%444444444444444444444444444444444444444444444444444

A particularly important issue, which has only recently received attention \cite{Villegas-2008}, is the determination of entropy in the Axelrod model. If the Shannon-Boltzmann-Gibbs configurational entropy ($S$), also called {\it cluster entropy} \cite{Tsang-2000}, is $S/N = - k_B \sum_{k = 1}^{N}\,p_k ln p_k$, where $k_B$ is the Boltzmann constant and $p_k = N_k /N$, where $N_k$ is the number of agents belonging to a domain of size $k$ formed in the absorbing state, then $p_k$ represents the probability that an agent is part of a domain of $k$ agents. Fig. 4 shows, for $F = 5$, $n_s = 50$ (or $n_s = 100$, as the case may be), and the four values of $N$ mentioned in this paper, the average value of $\overline S$, $\langle \overline S\rangle$, as a function of the number of traits $q$, where 
$\overline S = S /N k_B$. In this figure, and in subsequent ones, the vertical line separates the regions where $q < q_c$ (homogeneity, monoculturality) and $q > q_c$ (heterogeneity, multiculturalism).

%5555555555555555555555555555555555555555555555555555
\begin{figure}
	\centering
%	\fbox{\rule[-.5cm]{4cm}{4cm} \rule[-.5cm]{4cm}{0cm}}
%	\caption{Sample figure caption.}
% \includegraphics[width=\linewidth]
 \includegraphics[width=12.7cm]
    {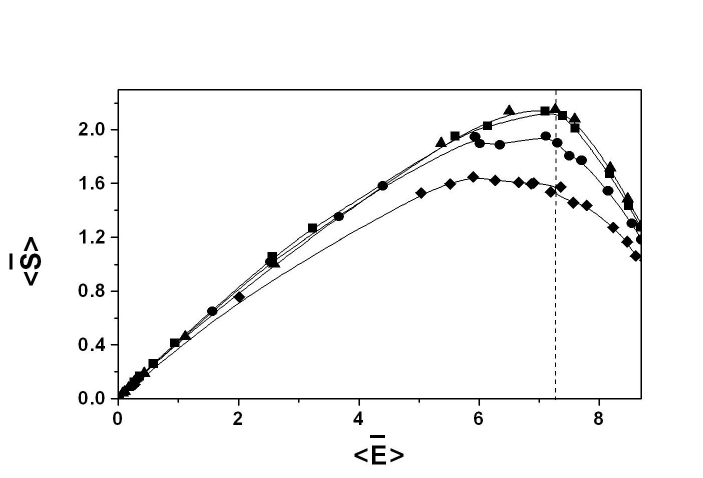}
 \caption{Average entropy reduced per agent $\langle \overline S\rangle$ as a function of the average energy reduced per agent
 $\langle \overline E\rangle$.}
 %.
 \label{figure5}
 %    }
%
\end{figure}
%5555555555555555555555555555555555555555555555555555

Most notable is the evidence of a maximum in the reduced entropy per agent $\overline S$ of the system, for a value of $q \cong 28$. This latter value is in the limit $N \rightarrow \infty$. Tsang (\cite{Tsang-2000}), Figure 7.13 of the cited reference, in the context of percolation theory, obtains a similar graph, inferring that the occupancy probability $p$ of a node in a square two-dimensional network is $p \sim 1/q$.

%6666666666666666666666666666666666666666666666666666
\begin{figure}
	\centering
%	\fbox{\rule[-.5cm]{4cm}{4cm} \rule[-.5cm]{4cm}{0cm}}
%	\caption{Sample figure caption.}
 %\includegraphics[width=\linewidth]
 \includegraphics[width=12.7cm]
    {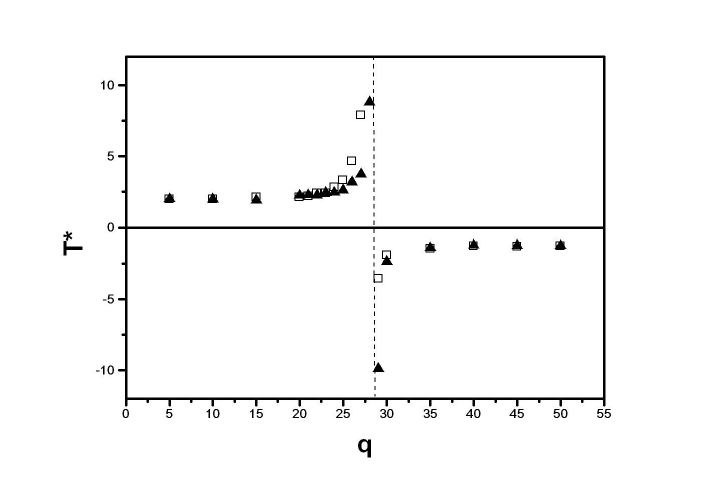}
 \caption{Temperature $T^*$ as a function of the traits $q$. The squares and triangles represent the results for $N = 40 \times 40$ and $60 \times 60$, respectively. The vertical line divides the regions with $T^* > 0$ and $T^* < 0$.}
 %.
 \label{figure6}
 %    }
%
\end{figure}
%6666666666666666666666666666666666666666666666666666

As we did in a recent paper \cite{Villegas-2008}, for $N = 20 \times 20$ and $40 \times 40$ agents, we then define temperature as $T^{*} = (\partial \overline E / \partial \overline S)$. Fig. 5 shows the average entropy reduced per agent $ \langle \overline S \rangle$ as a function of the average energy reduced per agent $\langle \overline E \rangle$, for the aforementioned values of $N$. Point by point, the slope $(\partial \overline S / \partial \overline E)^{-1}$ of each of the curves -for the corresponding $N$- is $1/T^{*}$. The maximum corresponds to $q = q_c \cong 28$, for the case when $N\rightarrow \infty$ and is the point where $T^{*} \rightarrow \pm\infty$. Therefore, the left side of the graph corresponds to states of the system with $T^{*} > 0$, while the right side corresponds to $T^{*} < 0$. In the figure, a vertical line separates both cases. This result is a product of the maximum entropy and the monotonic increase in energy as 
$q \rightarrow \infty$, as clearly observed in Figures 2 and 4.

%777777777777777777777777777777777777777777777777777
\begin{figure}
	\centering
%	\fbox{\rule[-.5cm]{4cm}{4cm} \rule[-.5cm]{4cm}{0cm}}
%	\caption{Sample figure caption.}
% \includegraphics[width=\linewidth]
 \includegraphics[width=12.7cm]
    {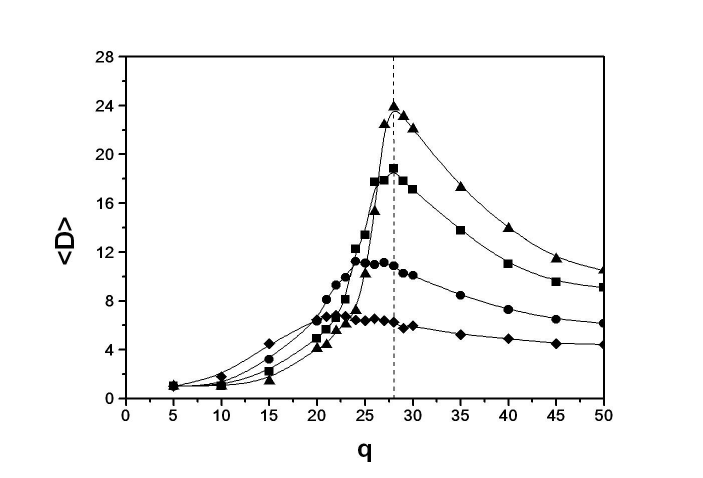}
 \caption{The average diversity $\langle D\rangle$ as a function of $q$.}
 %.
 \label{figure7}
 %    }
%
\end{figure}
%777777777777777777777777777777777777777777777777777

%888888888888888888888888888888888888888888888888888
\begin{figure}[h! tbp]
	\centering
%	\fbox{\rule[-.5cm]{4cm}{4cm} \rule[-.5cm]{4cm}{0cm}}
%	\caption{Sample figure caption.}
% \includegraphics[width=\linewidth]
 \includegraphics[width=12.7cm]
    {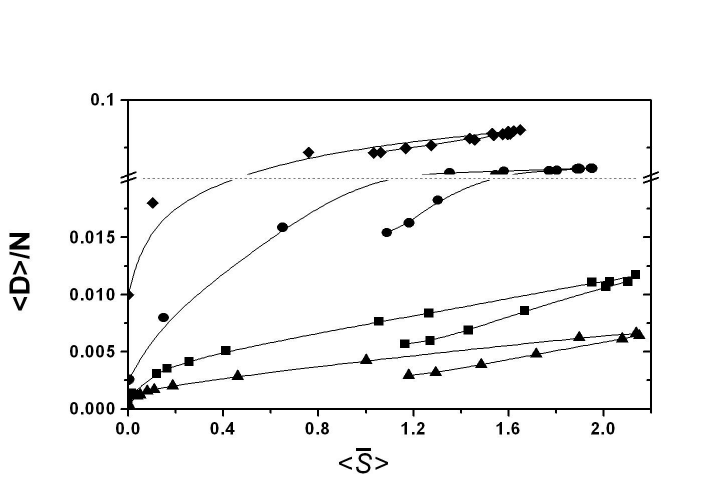}
 \caption{The average per-agent diversity $\langle D \rangle /N$ as a function of the average per-agent reduced entropy $\langle \overline S \rangle$.}
 %.
 \label{figure8}
 %    }
%
\end{figure}
%888888888888888888888888888888888888888888888888888

Figure 6 shows $T^*$ as a function of q, for two values: $N = 40 \times 40$ (squares) and $N = 60 \times 60$ (triangles). At the point where $T^* = (\partial \overline E / \partial \overline S) \rightarrow \pm \infty$, the moment where the order-disorder transition occurs (monocultural/multicultural), the sign of $T^*$ changes abruptly from positive to negative.
In this figure, a straight vertical line separates both cases. As can be seen in the aforementioned figure, this change is much better defined as $N \rightarrow \infty$.

Figure 7 shows the average diversity $\langle D \rangle$ as a function of $q$. As in the other figures, when $N \rightarrow \infty$ the results tend asymptotically toward limiting behavior. The vertical line, located at $q \cong q_c$, defines the two distinct regions we mentioned earlier: 

a) When $q < q_c$ , the system tends to be monocultural, and $D \rightarrow 1$, the minimum possible value.

b) When $q > q_c$ , the system tends to be multicultural, {\it but also} $D \rightarrow 1$, the minimum possible value. It is clear that at $q_c$, the maximum {\it cluster} diversity, $D_c$, is obtained.

It is evident that the diversity $D$ is linked to the entropy $\overline S$, because:

a) When $\overline S \rightarrow 0$, $D\rightarrow 1$; which occurs if $q \rightarrow 1$ and/or $q \rightarrow \infty$.

b) When $\overline S \rightarrow \overline S_c$, $D \rightarrow D_c$; which occurs if $q \rightarrow q_c$.

$D_c$ and $\overline S_c$ represent the diversity and reduced entropy per agent at the critical point, respectively.

Fig. 8 shows the average diversity per agent, $\langle D \rangle / N$, vs. the average reduced entropy per agent 
$\langle \overline S \rangle$, for the four
values of $N$ studied in this work. It is observed that $\langle D \rangle / N$ and $\langle \overline S \rangle$ increase and decrease monotonically, proportionally. 
As $N\rightarrow \infty$, the {\it hysteresis} region (so-called in this work because it recalls the asymmetry phenomena that appear in the magnetization and demagnetization processes of a ferromagnetic substance) is significantly reduced, with the respective curve reaching its maximum value at the critical point.

%xxxxxxxxxxxxxxxxxxxxxxxxxxxxxxxxxxxxxxxxxxxxxxxxx
\section{Dicussion of Results}
%xxxxxxxxxxxxxxxxxxxxxxxxxxxxxxxxxxxxxxxxxxxxxxxxx

As mentioned, $\langle M_{MAX}\rangle/N$ and $\langle D\rangle/N$ have traditionally been used as {\it order parameters}. The critical point, where the monocultural/multicultural transition occurs, is located where such curves show a change in concavity (inflection point). The results shown in this work are similar to those previously reported by \cite{Gonzalez-2006,Villegas-2008}, for $N = 40 \times 40$. When $N\rightarrow \infty$, which is approximately reached in a system with $N = 60 \times 60$, the critical point $q_c \cong 28$. It is clearly observed that $\langle M_{MAX}\rangle/N$ and $\langle D\rangle/N$, for $N = 60 \times 60$, already tend to exhibit asymptotic behavior and to be independent of the system size, as expected. 

Fig. 2 shows that an increase in $q$ leads to an increase in the system's energy $\overline E$, which is expected because it decreases the probability of overlaps in the cultural vectors of neighboring agents. This probability changes dramatically near the critical point, with a sudden change in the concavity of the $\overline E$ curve at $q_c$. If energy is a measure of system stability, then the homogeneous phase exhibits the greatest possible stability. 

In Fig. 4, we observe that the maximum in disorder is located precisely at the critical point ($\overline S = \overline S_c$). Below and above this point, disorder decreases. That is, the system becomes more ordered as $q \rightarrow 1$ and $q \rightarrow \infty$. However, for $q < q_c$, the system has a very small entropy, except in the region near the critical point, where it rapidly rises to its maximum value. For $q > 
q_c$ and $q \rightarrow \infty$, the system reorganizes, but much more slowly. 

Fig. 5 shows a combination of both trends: the monotonic increase in energy and the asymmetric "bell" shape of entropy as $q$ increases. 
Note that the derivative $(\partial \overline S / \partial \overline E)$ of the corresponding curve is nothing more $1/T^*$. At both extremes of this figure, when $\overline E \rightarrow 0$ and $\overline E \rightarrow 2F = 10$, $\overline S \rightarrow 0$. In such situations, the system tends toward maximum order, so $T^* \rightarrow 0$. However, as already pointed out in a previous work \cite{Villegas-2008}, there are actually two limits: the “left-hand” limit, when $T^*\rightarrow +0$,
and the “right-hand” limit, when $T^*\rightarrow -0$. At the maximum of the curves, however, either $T^*\rightarrow +\infty$ or 
$T^*\rightarrow -\infty$, depending on whether we are “coming” from the region $T^* > 0$ or from $T^* < 0$,
respectively. Fig. 6 shows this behavior.

The state where $T^*\rightarrow +0$, which has the lowest possible energy ($\overline E = 0$), must be a stable state. Conversely, when 
$T^*\rightarrow -0$ and $\overline E\rightarrow 2F = 10$ (the highest possible energy), the system must be {\it metastable}. Any perturbation can knock the system out of this precarious equilibrium. This could explain why multicultural states in Axelrod's model tend toward monoculturality when subjected to a set of small perturbations \cite{Klemm-2003b}.
%[Klemm et al. 2003b]. 

From this perspective, the fact that the states in the Axelrod model could be metastable when $q > q_c$ and stable when $q < q_c$ should not limit the application of equilibrium thermodynamic relations and the definition of state properties, such as energy, entropy, and temperature. In fact, strictly speaking, there is no real system that is exactly in equilibrium (i.e., executing a reversible process); since all real processes are finite in time; that is, irreversible \cite{Dickerson}. And it is the case that, as is widely known, experimental values of pressure, temperature, energy, etc., are reported daily in specialized papers and books in real physical or physicochemical systems, despite always corresponding, strictly speaking, to non-equilibrium states.

The two-dimensional Axelrod model can certainly be considered "quantum," since its possible energy states are discrete, enumerable, and bounded. It also clearly has a maximum possible value allowed for its energy, 
$\overline E = 2F$, and is therefore susceptible to negative temperatures $T^*$, as already mentioned. The sociocultural implications of the existence of absolute temperatures, both positive and negative, in Axelrod's model will not be analyzed in this paper and will be part of future research in this field.

The maximum reached in $\overline S$, when $q = q_c$ , also leads us to reconsider the equivalence between order $\leftrightarrow$ monoculturality ({\it globalization}) and disorder $\leftrightarrow$ multiculturalism ({\it polarization} or {\it cultural diversity}), mentioned in the literature \cite{Klemm-2003b, Gonzalez-2005}. 
Since $\overline S$ has a maximum at $q = q_c$ , and therefore $T^* \rightarrow \pm \infty$, it corresponds to a situation where the degrees of freedom of the system are maximum, and therefore also the diversity $D$. 

Tsang \cite{Tsang-2000} showed, in the aforementioned percolation study, that the maximum cluster diversity is located where entropy is also maximized, which in his case is at the percolation threshold. This makes sense, since $D$ and $\overline S$ maintain a nonlinear monotonic relationship (see Fig. 8). In our work, the maximum occurs at the critical point $q_c$. Above and below this threshold, diversity decreases and gradually tends to zero. It is then possible to establish that in Axelrod's model, maximum cultural diversity is found at $q = q_c$, and that when $q < q_c$ and/or $q > q_c$, the diversity of cultural domains must also tend toward its minimum value, as $q \rightarrow 1$ and $q \rightarrow \infty$, respectively. Therefore, {\it multiculturalism} is not synonymous with {\it cultural diversity}, and even more so, with {\it cultural richness}.

In this context, $q$ can be considered an "extrinsic" variable, that is, it recreates the material conditions available to the agents of this model society. When $q$ increases, from $1$ to $\infty$, we are actually increasing the material "opportunities" for access to broader cultural environments: forms and types of nutrition and food, entertainment, media and information, clothing, religion, political options, etc. This certainly leads, within the limits allowed by Axelrod's cultural model, to an increase in the cultural diversity displayed by this society. The maximum in this cultural diversity occurs precisely at the critical point, where entropy $\overline S$ is also maximized and temperature $T^*$ tends to infinity. However, if we continue increasing the degrees of freedom $q$ above $q_c$, the diversity of cultures begins to decrease, because they begin to repeat themselves massively. This repetition is expressed in the model as an increase in the number of clusters that have the same number of agents. Individuals no longer form diverse groups, but tend to fragment into many small groups, isolated from each other. This is particularly evident when $q \rightarrow\infty$. The latter seems to be what is observed in large, cosmopolitan cities: many opportunities to access goods and services lead to social isolation and the formation of many cultural groups of the same type, small in number, homogeneous, and highly compact. On the contrary, when $q\rightarrow 1$, we are talking about restricted sociocultural environments: for example, small villages and rural properties. As everyday experience would seem to indicate, here the human groups, culturally speaking, are few, highly homogeneous, but very numerous.

Ultimately, the monocultural/multicultural transition is not of the order-disorder type as we might subjectively think (when moving from few to many cultures, when $q\rightarrow\infty$), in analogy to phase transitions experienced by physical systems, but rather is the change from "one type of order" to "another type of order": few and repeated cultural groups with many agents (a homogeneous monocultural state) passing to many and repeated cultural groups with few agents (an inhomogeneous multicultural state).

\section{Conclusions}

In summary, Cultural Diversity ($D$) can be used as an {\it order parameter}, instead of $\langle M_{MAX} \rangle / N$ or $\langle N_D \rangle / N$, because it unambiguously locates the critical point where the monocultural/multicultural {\it phase transition} occurs for the Axelrod model. It is evident that at $q = q_c$, Cultural Diversity has a maximum, as does the reduced entropy per agent $\overline S$. It is shown that in the Axelrod model, as temperature tends to infinity, Diversity tends to its maximum value for each value of $N$. It is then shown that, analogously to what occurs in other percolation models in two-dimensional square networks already mentioned, the diversity of cultural domains in the Axelrod model is also maximum at the critical point. It is then found that cultural diversity tends to its minimum value ($D \rightarrow 1$ or $D/N \rightarrow 0$) when we are in monocultural or multicultural states. Therefore, {\it multiculturalism} is not synonymous with {\it cultural diversity}. Beyond the critical point, at high and low values of cultural traits $q$, the different cultures are repetitive, which is particularly surprising when $q \rightarrow\infty$. Seen through the lens of $D$ and $\overline S$, the monocultural/multicultural transition is of the {\it order-order} type, and not {\it order-disorder}, as commonly assumed in the literature. This opens up interesting perspectives regarding the interpretation of diversity and its application in fields such as economics, commerce, and communication products, among others.

\section*{Acknowledgments}

This work was partially supported by the ADG-CDCHT-ULA-C-09-05 program of the Scientific, Humanitarian, and Technological Development Council (CDCHT) of the University of Los Andes, Venezuela. J. Villegas-Febres thanks W. Olivares-Rivas and M. Cosenza for their suggestions and fruitful discussions.
%%%%%%%%%%%%%%%
%%%%%%%%%%%%%%%
%
\bibliographystyle{unsrtnat}

%xxxxxxxxxxxxxxxxxxxxxxxxxxxxxxxxxxxxxxxxxxxxxxxxxxxxxxxxxxxxxxx
%xxxxxxxxxxxxxxxxxxxxxxxxxxxxxxxxxxxxxxxxxxxxxxxxxxxxxxxxxxxxxxx
%xxxxxxxxxxxxxxxxxxxxxxxxxxxxxxxxxxxxxxxxxxxxxxxxxxxxxxxxxxxxxxx

\end{document}